\documentclass[prl,amsmath,notitlepage,twocolumn]{revtex4-2}
\usepackage{graphicx}
\usepackage{float}
\usepackage{epstopdf,cancel}
\usepackage{epsf,latexsym,bbm,euscript}
\usepackage{amssymb,amsmath}
\usepackage{mathtools} %used in this file for the command \coloneqq -> :=
\usepackage{times,graphics}
\usepackage{soul,xcolor}
\usepackage{mathtools}
\usepackage[normalem]{ulem}
\usepackage{setspace}

\def\6{{\langle}}
\def\9{{\rangle}}
\newcommand{\defeq}{\vcentcolon=}
\newcommand{\eqdef}{=\vcentcolon}

\newcommand{\be}{\begin{equation}}
	\newcommand{\ee}{\end{equation}}
\newcommand{\ba}{\begin{eqnarray}}
	\newcommand{\ea}{\end{eqnarray}}

\newcommand{\rin}{{\mathrm{in}}}

\newcommand{\fA}{{\mathfrak{A}}}

\usepackage{scalerel}
\usepackage{tikz}
\usetikzlibrary{svg.path}
\definecolor{orcidlogocol}{HTML}{A6CE39}
\tikzset{
	orcidlogo/.pic={
		\fill[orcidlogocol] svg{M256,128c0,70.7-57.3,128-128,128C57.3,256,0,198.7,0,128C0,57.3,57.3,0,128,0C198.7,0,256,57.3,256,128z};
		\fill[white] svg{M86.3,186.2H70.9V79.1h15.4v48.4V186.2z}
		svg{M108.9,79.1h41.6c39.6,0,57,28.3,57,53.6c0,27.5-21.5,53.6-56.8,53.6h-41.8V79.1z M124.3,172.4h24.5c34.9,0,42.9-26.5,42.9-39.7c0-21.5-13.7-39.7-43.7-39.7h-23.7V172.4z}
		svg{M88.7,56.8c0,5.5-4.5,10.1-10.1,10.1c-5.6,0-10.1-4.6-10.1-10.1c0-5.6,4.5-10.1,10.1-10.1C84.2,46.7,88.7,51.3,88.7,56.8z};
	}
}

\newcommand\orcidlink[1]{\href{https://orcid.org/#1}{\mbox{\scalerel*{
				\begin{tikzpicture}[yscale=-1,transform shape]
					\pic{orcidlogo};
			\end{tikzpicture}}{X}}}}

\def\etal{\textit{et al.}}

\def\1{{{\mathbbm 1}}}

\def\pad{{\partial}}

\def\sg{\textsl{g}}

\def\cO{\mathcal{O}}

\def\mM{\mathrm{M}}
\def\mP{m_\mathrm{Pl}}

\def\tT{t_\mathrm{T}}

\def\vKo{\mathbbm{k}}

\usepackage{url,hyperref}
\hypersetup{colorlinks,linkcolor={blue!55!black},citecolor={red!45!black},urlcolor={blue!45!black},breaklinks=true}

\begin{document}

	\title{Models of cosmological black holes}
	\author {Pravin K. Dahal\,\orcidlink{0000-0003-3082-7853}}
	\email{pravin-kumar.dahal@hdr.mq.edu.au}
	
	\author{Swayamsiddha Maharana\,\orcidlink{0009-0004-6006-8637}}
	\email{swayamsiddha.maharana@hdr.mq.edu.au}
	
	\author{Fil Simovic\,\orcidlink{0000-0003-1736-8779}}
	\email{fil.simovic@mq.edu.au}
	
	\author{Ioannis Soranidis\,\orcidlink{0000-0002-8652-9874}}
	\email{ioannis.soranidis@hdr.mq.edu.au}
	
	\author{Daniel R.\ Terno\,\orcidlink{0000-0002-0779-0100}}
	\email{daniel.terno@mq.edu.au}
	
	\affiliation{School of Mathematical and Physical Sciences, Macquarie University, NSW 2109, Australia}

	\begin{abstract}
		\vspace*{1mm}
		We study various aspects of modeling astrophysical black holes using the recently introduced semiclassical formalism of physical black holes (PBHs). This approach is based on the minimal requirements of observability and regularity of the horizons. We demonstrate that PBHs do not directly couple to the cosmological background in the current epoch, and their equation of state renders them unsuitable for describing dark energy. Utilizing their properties  for analysis of more exotic models,   we present a consistent semiclassical scenario for a black-to-white hole bounce and identify obstacles to the transformation from a black hole horizon to a wormhole mouth.

		%\medskip

	\end{abstract}
	
	\maketitle
	
	\textit{Introduction.}--- %Astrophysical black holes (ABHs)  drive some of the most energetic  phenomena in the Universe.
	More than a hundred astrophysical black holes (ABHs) — dark, massive, ultra-compact objects — have been identified \cite{LIGO:21,LIGO:23,EHT:19,EHT:22}. % Accelerated detection and increasing precision in parameter %estimation highlight the lack of consensus regarding their nature. 
All observations are well-described by the classical Kerr solution to Einstein-Hilbert gravity \cite{LIGO:23,EHT:22,CP:19}. This consistency is not yet a demonstration of key  {black hole} features, such as light trapping. Moreover,   {a literal} identification of ABHs with classical black hole solutions requires addressing singularities and unresolved semiclassical issues. This tension motivates development of the alternative ABH models, many of which are observationally viable \cite{bh-plan:19,CP:19}.
	
	%More than a hundred astrophysical black holes (ABHs) — dark, massive, ultra-compact objects — have already been identified \cite{LIGO:21,LIGO:23,EHT:19,EHT:22}.  The accelerated detection and the increasing precision %with which their parameters are determined bring into sharp focus the question of their nature. All observations are thus far well described by the classical Kerr solution to Einstein-Hilbert gravity %\cite{LIGO:23,EHT:22,CP:19}. However, this consistency is not yet a demonstration of key features, such as the trapping of light, while their full acceptance requires addressing singularities and unresolved %semiclassical issues. These issues have motivated the introduction of alternative ABH models, many of which remain observationally viable \cite{bh-plan:19,CP:19}.
	
	A rough classification scheme differentiates ABH models with horizons from their horizonless counterparts. The event horizon, a null surface that causally disconnects the black hole's interior from the outside world \cite{HE:73,FN:98}, is the defining feature of a mathematical black hole (MBH) \cite{F:14}. Physical black holes (PBHs) \cite{F:14} are trapped spacetime regions, potentially transient and not always overlapping with MBHs \cite{MMT:22}. In contrast, various models describe ABHs without introducing horizons \cite{CP:19}. Distinguishing among ABH models remains one of the central problems in black hole physics \cite{bh-plan:19}.
	
	%A rough  classification scheme differentiates between ABH models with horizons and their horizonless counterparts. The event horizon, a null surface that
	%causally disconnects the black hole interior from the outside world \cite{HE:73,FN:98}  is the defining feature of a mathematical black hole (MBH) \cite{F:14}. Physical black holes (PBHs) \cite{F:14}, on the other %hand, are trapped spacetime regions, possibly transient and not necessarily overlapping with MBHs \cite{MMT:22}. On the other hand, a variety of models  purport to describe  ABHs without introducing a horizon %\cite{CP:19}.  One of the central  problems in black hole physics involves distinguishing between the variety of ABH models that have been proposed \cite{bh-plan:19}.

	In fact, we should distinguish three possibilities \cite{MMT:22,M:23}. From a distant observer's point of view, gravitational collapse beyond neutron star density can result in either:
	(i) Ongoing collapse, with a horizon as an asymptotic ($t \to \infty$) concept. Under broad conditions (the energy conditions discussed below), MBHs are gravitational collapse's asymptotic states \cite{HE:73,FN:98};
	(ii) Formation of a transient or stable object, where a  {suitably-defined}   deviation from $r_\sg$ reaches a minimum $\varepsilon>0$ \cite{CP:19}. Black hole mimickers  fall into this category;
	(iii) Formation of an apparent horizon in finite $t_\text{f}$. PBHs, subject to this requirement, fall into this category \cite{MMT:22,DSST:23}.
	The same alternatives arise in  mergers.

	%In fact, we should distinguish between three possibilities \cite{MMT:22,M:23}. From the point of view of a distant observer the gravitational collapse beyond the density of a neutron star can have three possible %outcomes:
	%	(i) Perpetual collapse,  with a   horizon being an asymptotic ($t \to \infty$) concept. Under quite broad conditions (the energy conditions that we discuss below), MBHs are the asymptotic states of gravitational %collapse \cite{HE:73,FN:98}; (ii)  Formation of a transient or a stable  object, where a suitable   parameter \cite{CP:19} that measures the deviation from $r_\sg$ reaches a minimal value. Black hole mimickers --- %exotic compact objects  --- belong to this category; (iii) Formation of an apparent horizon     in finite  $t_\text{f}$.     PBHs, subject to this additional requirement  belong to this category \cite{MMT:22,DSST:23}. %The same alternatives arise in a mergers.
	
	Here, we  investigate the consequences of accepting  option (iii). After describing PBHs, we analyse their direct coupling with cosmological dynamics, identify a viable black-to-white hole transition scenarios, and present the obstacles in forming traversable wormholes. It is noteworthy that, like most  conceptual and observational black hole discussions,   our analysis is framed within the semiclassical framework. Thus, our conclusions serve both as a list of PBH features, and as indicators of potential breakdown of semiclassical gravity on a macroscopic scale.

	\textit{Semiclassical  black and white holes}. --- Our formalism is based on two  premises. First,  {the} spacetime geometry can be described using a metric $\sg_{\mu\nu}$ with classical concepts of horizons, trajectories,  and the equations that describe them remaining valid.  Second, we assume the metric $\sg_{\mu\nu}$  is a solution to the  {semiclassical} Einstein equations \cite{FW:96,cK:12,HV:20}
	\be
	G_{\mu\nu}=8\pi\6\hat{T}_{\mu\nu}\9_\omega \ , \label{ee}
	\ee
	where the left hand side is the Einstein tensor $G_{\mu\nu}=R_{\mu\nu}-\tfrac{1}{2}R\sg_{\mu\nu}$, and the right hand side is the effective energy-momentum tensor (EMT),   $\6\hat{T}_{\mu\nu}\9_\omega\equiv T_{\mu\nu}$. It includes the renormalized expectation value of all matter fields,  higher-order terms arising from its regularisation, and possible contributions arising from modifications to  {the} Einstein-Hilbert gravity or a cosmological constant $\Lambda$. In the subsequent analysis we do not use any specific property of the state $\omega$ and do not separate the matter EMT into the collapsing matter and (perturbatively-obtained) quantum excitations.
	
	In discussing PBH properties \cite{MMT:22}, we apply the weakest form of the  cosmic censorship and require absence of scalar curvature singularities at the apparent horizon \cite{HE:73}. Unlike the teleological event horizon, the apparent horizon is in principle observable \cite{mV:14}.  {We express this observability by requiring a finite  formation time according to the  clocks of distant observers} \cite{MsMT:22,MMT:22}.

	Most of our discussion  is restricted to spherical symmetry, where the two above assumptions lead to the exhaustive classification of allowed metrics and  to clarification of some key features of the black hole formation and the near-horizon geometry. We will also use Kerr--Vaidya black holes as the simplest example of  axially-symmetric dynamical  models \cite{DT:20,DMST:23}.
	
	A general spherically symmetric metric in  Schwarzschild coordinates ($r$ is the areal radius) is given by
	\be
	ds^2=-e^{2h(t,r)}f(t,r)dt^2+f(t,r)^{-1}dr^2+r^2d\Omega_2, \label{sgenm}
	\ee
	while using the advanced null coordinate $v$ results in the form
	\be
	ds^2=-e^{2h_+(v,r)}f_+(v,r)dv^2+2e^{h_+(v,r)}dvdr+r^2d\Omega_2\ . \label{m:vr}
	\ee
	The function $f$ is coordinate-independent, i.e. $f(t,r)\equiv f_+\big(v(t,r),r\big)$ and in what follows we omit the subscript. It is conveniently represented via the Misner--Sharp--Hernandez (MSH) mass $M_\mM\equiv C_\mM/2$ \cite{vF:15} as
	\be
	f=1-\frac{C_\mM(t,r)}{r}=1-\frac{C_+^\mM(v,r)}{r}=\pad_\mu r \pad^\mu r\ ,
	\ee
	The functions $h$ and $h_+$ play the role of integrating factors in the coordinate transformation
	\be
	dt=e^{-h}(e^{h_+}dv-f^{-1}dr)\ . \label{trvr-transformation}
	\ee
	For example, the Schwarzschild metric corresponds to $h\equiv 0$, $C_\mM\equiv r_\sg=\mathrm{const}$, and $v=t+r_*$, where $r_*$ is the tortoise coordinate \cite{FN:98,HE:73}, while the de Sitter metric in the static patch is given by $h\equiv 0$, $C_\mM=H^2 r^3$, where $H$ is the constant Hubble parameter \cite{vF:15,FGF:21}.

	\begin{figure}[htbp]
	\vspace{-25mm}
		\hspace*{-1.5cm}
		\includegraphics[width=1\linewidth]{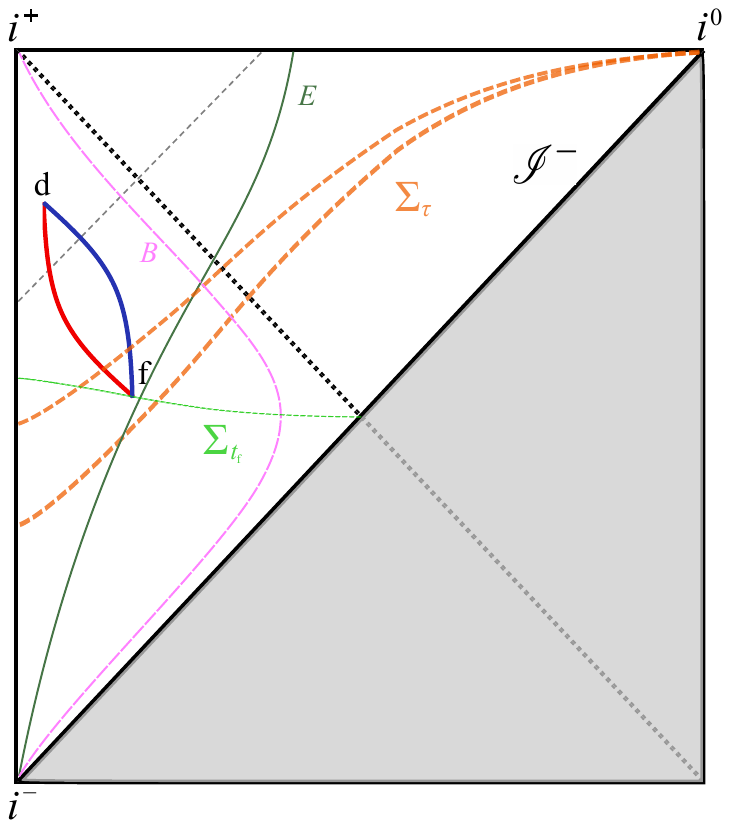}
\vspace{-40mm}
		
		\caption{Schematic Carter–Penrose diagram depicting the formation and evaporation of a regular black hole in a de Sitter spacetime \cite{DSST:23}.   Past and future timelike infinity are labelled by $i^-$ and $i^+$, respectively. Spacelike infinity is labelled by $i^0$.
			Dashed grey lines correspond to outgoing radial null geodesics. The trajectory of a distant observer  is indicated in pink and labelled $B$. The points $\mathrm{f}$ and $\mathrm{d}$ represent the two-spheres of formation and disappearance of the trapped region. The equal (Schwarzschild) time hypersurface $\Sigma_{t_{\mathrm{f}}}$ is shown as a dashed light green line.  The outer (blue) and inner (dark red) components of the black hole apparent horizon (timelike membranes) are indicated according to the invariant definition \cite{vF:15,MMT:22}. The invariantly-defined components of the apparent horizon correspond   the largest and smallest root of $f=0$ whether $t$, $v$ or $u$ is used as the evolution parameter. The solid black line connecting $i^{-}$ and $i^{0}$ represents the cosmological event horizon for an observer at $r=0$. Static coordinates cover only the left quadrant, with the dotted diagonal line representing the particle horizon. Components of the black hole apparent horizon correspond to the largest and smallest roots of $f=0$ (not including the cosmological horizon). The orange dashed lines $\Sigma_{\tau}$ indicate hypersurfaces of constant  {cosmological time $\tau$}. The trajectory of an asymptotically comoving observer  ($\chi=\mathrm{const}$)  is marked by the dark green line and labelled by the initial $E$.}
		\label{fig:rbh-dS}
	\end{figure}
	
	%\vspace{-8mm}
	
	In a cosmological setting  (such as depicted on Fig.~\ref{fig:rbh-dS}), we assume that a separation of scales exists between geometric features associated with the black hole and those of the large-scale universe \cite{DSST:23}. In this case, the apparent horizons of the PBH is given by the real roots of $f(t,r)=0$ in the near-region ~\cite{MMT:22}. The largest such root is the Schwarzschild radius $r_\sg$ that represents the location of the outer black hole horizon. Invariance of the MSH mass implies
	\be\label{msh}
	r_\sg(t)=r_+\big(v(t,r_\sg(t)\big)\ ,
	\ee
	where $r_+$ is the equivalent root of $f(v,r)$. Unlike the globally defined event horizon, the   apparent horizon is foliation-dependent. However, it is invariantly defined in all foliations that respect spherical symmetry \cite{FEFHM:17} which will be used in the following.

	 In preparation for the study of a PBH as an inhomogeneity on the cosmological background, described by $\Lambda$--Cold Dark Matter \cite{DS:21},  we set
	\be
	T_{\mu\nu}=T_{\mu\nu}^\mathrm{mat}-\Lambda \sg_{\mu\nu}/8\pi\ , \label{eq:Gtt-L}
	\ee
	separating the EMT into the matter and the cosmological constant  parts, respectively. Both the Einstein equations and the regularity requirements are most conveniently written if we introduce the effective EMT components
	\begin{align}
		\tau{_t} \defeq &e^{-2h} {T}_{tt}=\tau_t^\mathrm{mat}+\Lambda f/8\pi\ , \label{split1} \\
		\qquad {\tau}{^r} \defeq & T^{rr}=\tau^r_\mathrm{mat}-\Lambda f/8\pi\ , \label{split2} \\
		\tau {_t^r} \defeq & e^{-h}  {T}{_t^r}=\tau_t^{r\mathrm{mat}}  . \label{eq:mtgEMTdecomp}
	\end{align}
	The Einstein equations for the components $G_{tt}$, ${G}{_t^r}$, and ${G}^{rr}$ are then, respectively
	\begin{align}
		&\partial_r C_\mM = 8 \pi r^2  {\tau}{_t}^\mathrm{mat} / f+\Lambda r^2\ , \label{eq:Gtt} \\
		&\partial_t C_\mM = 8 \pi r^2 e^h  \tau_t^{r\,\mathrm{mat}}\ , \label{eq:Gtr} \\
		&\partial_r h = 4 \pi r \left(  \tau_t^\mathrm{mat} +  \tau^r_\mathrm{mat} \right) / f^2\ . \label{eq:Grr}
	\end{align}
	Hence the most general metric that describes the cosmological embedding of a spherical inhomogeneity into spatially flat de Sitter spacetime is given by Eq.~\eqref{sgenm} with
	\be
	C_\mM=C(t,r)+H^2r^3 \ ,
	\ee
	where $C$ (formally) satisfies the Einstein equations without the cosmological constant and $H=\sqrt{\Lambda/3}$.
	
	To enforce the regularity condition it  is  sufficient to ensure that only {the} curvature scalars $R$ and $R_{\mu\nu}R^{\mu\nu}$ are finite \cite{MMT:22}. Moreover, the cosmological constant drops out of the key equations, and the analysis is performed along the usual lines of the self-consistent approach \cite{MMT:22} (see Supplementary Material). As a result, $C$ and $C_\mM$ have the same structure \cite{DSST:23} as $r\to r_\sg$. There are two admissible classes of the near-horizon solutions that are distinguished by the scaling of the effective EMT components as $f^k$, $k=0,1$ when $r\to r_\sg$.
	
	The three components  scale for the generic  ($k=0$) solution as
	\be
	\tau_t,\tau^r \to -\Upsilon^2, \qquad \tau_t^r\to\pm \Upsilon^2\ ,
	\ee
	 for some $\Upsilon(t)$.
	The two metric functions are \be
	C= r_\sg-4\sqrt{\pi}r_\sg^{3/2}\Upsilon\sqrt{x}+\cO(x), \qquad h=-\frac{1}{2} \ln{\frac{x}{\xi}}+\cO(\sqrt{x}), \label{k0met}
	\ee
	where  $x\defeq r-r_\sg(t)$, and the function  $\xi(t)$ is determined by the  choice of the time variable.
	
	Consistency of the Einstein equations requires that
	\be
	\frac{d r_\sg}{dt} {\equiv r'_\sg} =\pm4\Upsilon\sqrt{\pi r_\sg\xi},      \label{lumin}
	\ee
	where the plus (minus) sign corresponds to expansion (contraction) of the  Schwarzschild radius.
	The case of $r_\sg'<0$ is most conveniently described using the advanced null coordinate $v$. Evaluation of the expansions of null geodesic congruences \cite{HE:73,vF:15} identifies the domain $f<0$ as a trapped region, and thus a PBH. The case of $r_\sg'>0$ is most conveniently described using the retarded null coordinate $u$. In this case the domain $f<0$ is the anti-trapped region, and we call the resulting entity a physical white hole (PWH) \cite{MMT:22}.
	
	The function $\Upsilon^2(t)>0$ determines the energy density  at the apparent horizon,  and the higher-order terms are matched with higher-order terms in the EMT expansion \cite{MMT:22}. Both solutions violate the null energy condition (NEC) \cite{HE:73,KS:20}, i.e. there are null vectors $k^\mu$ such that $T_{\mu\nu}k^\mu k^\nu<0$.  This is consistent with the result \cite{HE:73,FN:98} that the apparent horizon is not `visible' to a distant observer unless the NEC is violated.

	In both cases, the hypersurfaces $r=r_\sg$ are timelike. Therefore, both null (as indicated by Eq.~\eqref{lumin}) and sufficiently fast massive test particles can enter a PBH in finite time $t$. The same applies to a PWH, although the analysis is somewhat more complex \cite{DMST:23}.
	
	Regular black holes (RBHs) are non-singular PBHs. As such, they must have an inner horizon, which can be either timelike or null \cite{H:06}.   Analysis that is similar to the above shows that it is then a timelike hypersurface and the NEC is satisfied in its vicinity \cite{DMT:22,MS:23}.   The  horizons  do not joint smoothly, enabling the foliation-independent characterisation of the inner and outer segments \cite{DSST:23,MS:23}.

	Two properties are crucial for the following. First, after formation of a spherically symmetric PBH, its outer apparent horizon can only contract, meaning the PBH mass decreases. Conversely, a PWH can only grow. In axially symmetric models like Kerr--Vaidya black holes, the situation is more complex. Expansion of the apparent horizon in $(v,r)$ coordinates and contraction of the anti-trapping horizon in $(u,r)$ coordinates are not obviously excluded. Nevertheless, Ref.~\cite{DMST:23} identifies two issues. During a black hole's growth, signals from the apparent horizon cannot cross the hypersurface $r=M(v)+\sqrt{M^2(v)-a^2}$, which   for expanding black holes is  spacelike and screens the apparent horizon. Some white hole spacetimes with decreasing $M(u)$ are timelike geodesically incomplete.
	
	Second, the equation of state near the apparent or anti-trapping horizon is $\rho=p$. For small angular momentum values, this applies to Kerr--Vaidya black holes. At the outer horizon, both energy density and pressure are negative, while at the inner horizon, they are positive. Therefore, the PBH equation of state is more exotic than the de Sitter vacuum $\rho=-p>0$   that is used as a core in a variety of models \cite{CP:19}.

	%Two properties are particularly important for the following. First, once formed, the outer apparent horizon of a spherically symmetric PBH can only shrink, i.e.  its mass  can only decrease. On the other hand, a PWH %can only grow.  The situation is more involved  for the simplest axially symmetric models with variable mass --- Kerr--Vaidya black holes. Expansion of the apparent horizon  in regular metrics in $(v,r)$ coordinates %and shrinking of the anti-trapping horizon in regular metrics  in $(u,r)$ coordinates are not excluded a priori. While we cannot rule out a real-valued coordinate transformation that can establish a finite time $t$ of %formation, two problematic properties were established in Ref.~\cite{DMST:23}. During   growth of a black hole  with the mass $M(v)$  and the angular momentum to mass ratio $a$, no signals   from   the apparent horizon %can escape the hypersurface $r=M(v)+\sqrt{M^2(v)-a^2}$. At least some spacetimes of white holes with  decreasing $M(u)$  are timelike geodesically incomplete.
	
	% Second, the equation of state in the vicinity of the apparent horizon or the anti-trapping horizon is $\rho=+p$ (for small values of the angular momentum, this is approximately true also for the Kerr--Vaidya black %holes).  At the outer horizon, both the energy density and pressure are negative, while  at the inner horizon both density and pressure are positive. Thus the PBH equation of state is in some sense even more exotic %than the de Sitter vacuum $\rho=-p>0$ of some exotic models, such as gravastars \cite{CP:19,M:23}.
	
	\textit{Cosmological coupling of black holes}.---      Within
	sub-percent precision  the Universe is described at cosmological
	scales by spatially flat Friedmann--Lema\^{i}tre--Robertson--Walker (FLRW) metric \cite{DS:21,planck6}
	\be
	ds^2=-d\tau^2+a^2(\tau)(d\chi^2+\chi^2 d\Omega_2) \ ,
	\ee
	where $\chi$ is the comoving distance, $\tau$ the cosmic standard time, and $a(\tau)$ the scale factor. The Hubble parameter $H\defeq \dot a/a$.  In the present $\Lambda$-dominated epoch this geometry is well-approximated by the spatial flat de Sitter metric with $a(\tau)=H^{-1}\exp(H\tau)$.
	
	Smaller scale inhomogenuities, including embedding of the black holes in this cosmological background \cite{vF:15,FGF:21,AM:11} are active research subjects. In particular, the question whether black holes are affected by the large-scale dynamics of the cosmological background \cite{CMPS:23} was recently revisited. It was proposed in Ref.~\cite{CZNFT:21} that the black hole masses vary with the scale factor according to
	\be
	M(a)=M(a_\rin)\left(\frac{a}{a_\rin}\right)^q \ , \label{qLaw}
	\ee
	where $a_\mathrm{in}<a$  is the scale factor at which the object becomes
	cosmologically coupled, and $a$ is its current value.
	
	Ref.~\cite{FCZT:23} then reported a value
	$q=3.11^{+1.19}_{-1.33}$, and a strong confidence in excluding $q=0$.  This growth is immediately consistent with a cosmological vacuum equation of state $\rho=-p$ and thus black holes are a source of the dark energy.
	These conclusions, however, were criticised on both observational \cite{R:23,JWST:23,AE:23,CSPBM:23}  and theoretical \cite{A:23,CMPS:23,CSPBM:23,P:23,GV:23} grounds.
	
	The known explicit embedding models (Schwarzschild--Kottler--de Sitter, McVittie \cite{vF:15,FGF:21} and Kerr--de Sitter \cite{AM:11} show no cosmological coupling \cite{CMPS:23,GV:23}. Analysis of general scenarios typically focuses on   coupling of the cosmological background to local, spherically symmetric objects, since the black hole angular momentum effects are expected to be negligible on cosmological scales \cite{CMPS:23}.  Despite ambiguity of black hole definitions (MBH and PBH are just two of them, see Ref.~\cite{eC:19} for the discussion), in spherical symmetry a  {quasi}-local MSH mass, that coincides with the Hawking--{Hayward} mass, is the preferred quantity \cite{CMPS:23,CSPBM:23}. Moreover, it is directly tied to the notion of a PBH, as $M_\mM\equiv r_\sg/2$. Hence Eq.~\eqref{qLaw} implies
	\be
	\frac{dr_\sg}{d a}=\frac{\dot r_\sg}{\dot a}=q\frac{r_\sg}{a},
	\ee
	where derivatives both of the Schwarzschild radius and the scale factor are taken with respect to the cosmological time.
	
	As a result
	\be
	\frac{dr_\sg}{d a}=\frac{\dot r_\sg}{a H}\approx\frac{r'_\sg}{r_\sg \, H}\frac{r_\sg}{a},
	\ee
	where we assumed that the distant observer is in the asymptotically de Sitter region but still far from the cosmological horizon,  {i.e.} $r_\sg\ll r\ll H^{-1}$  {and} in this case $dt\approx d\tau$.  Eq.~\eqref{lumin} implies that $dr_\sg/da$ is then negative. So long as the evaporation does not dramatically deviate from the  Hawking--Page form \cite{FN:98} $r_\sg'\propto r_\sg^{-2}$,    at the current epoch $1\gg|q|\sim 0$. On the other hand, for the primordial black holes  at the end of inflation and/or early radiation-dominated epochs \cite{DS:21,K:10,Y:22} (that are treated as PBHs), the cosmological coupling   can be strong.
	
	\textit{Black-to-white hole bounce}.---
	The idea that quantum effects can prevent black hole singularity from forming can be realised in  several ways. For example, reformulationg the one-loop corrections as effectively modifying the Langrangian to fourth order, and collapse of a thin null shell results in transient, even if long-lived, trapped region \cite{FV:81}. Loop quantum gravity (LQG) corrections \cite{AB:05,BGMS:05}, similarly to their effect in quantum cosmology, modify the collapse (for  {example}, by introducing additional terms the classical Oppenheimer--Snyder collapse \cite{BGMS:05,M:17} leading to the singularity avoidance and transient trapped region.

	%The applicability of white hole solutions in general relativity to physically realizable scenarios has historically been met with scepticism.
	Over the past decade, there has been a resurgence in interest in white hole solutions, once considered less physically relevant predictions of general relativity \cite{HE:73,FN:98,vF:15}. Black-to-white hole transition scenarios \cite{M:17,A:20,HR:15,R:17,BCDHR:18,MR:19,RV:18,BBMU:20,RFLPV:22,HRS:23} have elevated their status. Two versions are discussed \cite{MR:19,BBMU:20,HRS:23}:  {``fireworks"} \cite{HR:15} and {``Planck stars"} \cite{R:17}, both modeled as tunneling processes with differing characteristic times $t_\mathrm{T}$. In the former, tunneling probability is  {of the order} $\exp{(-M^2/\mP^2)}$, thus $\tT\sim M^2$, with $M= r_\sg/2$ as the pre-transition black hole mass and $\mP$ the Planck mass. Since the Hawking–Page evaporation time scales as $r_\sg^3$, black and white holes are macroscopic, and Hawking radiation effects are negligible. Recently, an LQG-inspired modification of the Oppenheimer-Snyder model was described \cite{HRS:23}, featuring a single asymptotic region with a complete, singularity-free Lorentzian metric that satisfies Einstein equations up to the tunneling region. In the latter scenario, $\tT\sim M^3$, with tunneling occurring after the black hole shrinks to about $\mP$. This potentially long-lived Planck-scale remnant is suggested as a dark matter component \cite{RV:18}.

	%	Over the past decade  there has been a resurgence of interest in white hole solutions,  previously relegated to the less physically relevant results of general relativity \cite{HE:73,FN:98,vF:15}. The black-to-white %hole transition scenarios \cite{M:17,A:20,HR:15,R:17,BCDHR:18,MR:19,RV:18,BBMU:20,RFLPV:22,HRS:23}, brought them again as objects of interest. Two versions are discussed in the literature \cite{MR:19,BBMU:20,HRS:23}: %the so-called fireworks \cite{HR:15} and Planck stars \cite{R:17}, both modelled as a tunneling process but with different characteristic times $t_\mathrm{T}$. In the former scenario the tunneling probability is of the %order $\exp{(-M^2/\mP^2)}$ and thus $\tT\sim M^2$, where $M=2r_\sg$ is the black hole mass before the transition and $\mP$ is the Planck mass.  As the evaporation time according to the Hawking--Page law scales as %$r_\sg^3$, the black and white holes are macroscopic and the effects of Hawking radiation are neglected.  Recently \cite{HRS:23} an LQG-inspired modification of the Oppenheimer-Snyder model with a single asymptotic %region with a full singularity-free Lorentzian metric satisfying the Einstein equations up to the tunnelling region has been described. In the latter case $\tT\sim M^3$ and the tunneling occurs after the black hole %shrinks to the order of $\mP$. The potentially long-lived Planck-scale remnant is proposed as a possible  component of dark matter \cite{RV:18}

	\begin{figure}[htbp]
		\includegraphics[width=0.7\linewidth]{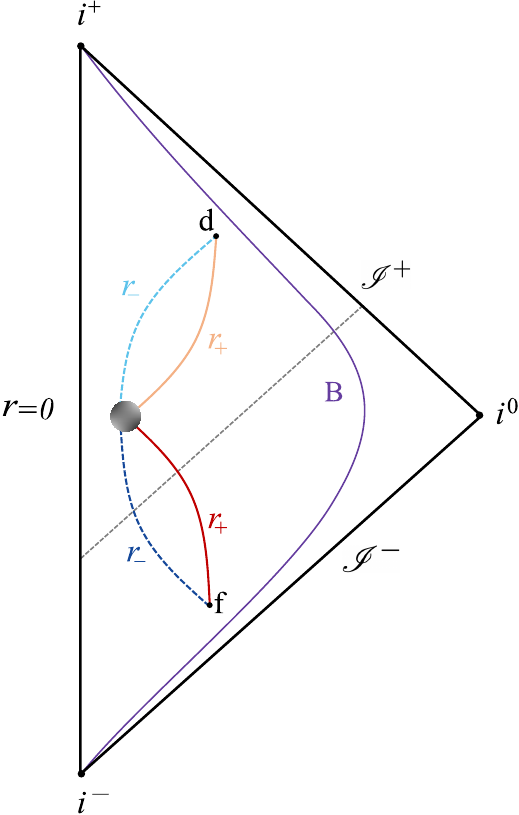}
		\caption{A schematic Carter--Penrose diagram for the formation of a black hole, transition to a white hole, and subsequent disappearance.  {Semiclassical} violation of the NEC implies timelike apparent trapping and anti-trapping horizons and a finite lifetime of the resulting white hole. }
		\label{fig:example}
	\end{figure}
%	\vspace{-7mm}
	
	Regardless of scenario specifics, it is agreed that the transition occurs in a regime where semiclassical physics breaks down \cite{M:17,A:20}. However, significant parts of both black and white hole domains are described semiclassically. Depending on their specifics, bounce scenarios may involve spacelike and timelike horizon segments, as well as a constant final white hole mass. Our approach limits these possibilities. Fig.~\eqref{fig:example} shows a modified semiclassical Penrose diagram of the black-to-white hole transition with a single asymptotic region based on this model. Within the validity domain of semiclassical gravity, objects before and after the bounce are PBH and PWH, respectively. While agnostic about the actual bounce mechanism, the timelike nature of PBH's inner and outer apparent horizons determines the pre-bounce structure. As only  {$r'_\sg>0$} white hole solutions are allowed, growth termination requires closing the anti-trapped region, leading to a massive object. Its subsequent evolution is undefined in our model, but it is momentarily singularity-free and contains neither trapped nor anti-trapped regions.

	%Regardless of the scenario specifics it is agreed  that the transition that occurs in the regime where the semiclassical physics breaks down \cite{M:17,A:20}. Nevertheless, significant partd of both black and white %hole domains are described semiclassically. Depending on their specifics the bounce scenarios may involve both spacelike and timelike segments of the horizons, as well as constant final white hole mass. Our approach %restricts these possibilities.   Fig.~\eqref{fig:example}  depicts a modified semi-classical Penrose diagram of the black-to-white hole transition with a single asymptotic region based on this model. Within domain of %validity of the semiclassical gravity the objects before and after the bounce are PBH and PWH, respectively.   While we remain agnostic  regarding the actual bounce mechanism, the timelike nature of the inner and outer %apparent horizons of a PBH determines the pre-bounce structure. As only $r_\sg>0$ white hole solutions are allowed,  the only way to terminate its growth is via closure of the anti-trapped region, leading to a massive %object whose subsequent evolution is not prescribed by our considerations, but its (momentarily, at least) singularity free and contains netierh trapped nor anti-trapped regions.

	\textit{Wormholes}.---
	A traversable wormhole (TWH) is a hypothetical shortcut in spacetime that connects two distant regions of our Universe or two separate Universes \cite{V:96,K:18}.  As the NEC violation   \cite{HV:98,K:18} and closed timelike curves are their unavoidable attributes, their primary use was in the realm of science fiction. However, in the last two decades they are considered as another class of ABH models \cite{CP:19,DS:07,CFP:16,SV:19,B:22} their potential observational signatures are investigated \cite{CP:19,CFP:16}.
	
	The original   wormhole solutions were characterized using the embedding diagrams and explicit description of the
	two spatial sheets that are connected at the wormhole’s throat. Wormhole metrics are often written using  the Schwarzschild coordinates on each sheet,
	\be
	ds^2=-e^\Phi dt^2+\frac{dr^2}{1-C/r} +r^2 d\Omega_2, \label{emtM}
	\ee
	where $\Phi$ is a regular function and $r\geqslant b_0>0$, where $b_0$ is the (largest) root of $C(t,r)=r$ \cite{K:18} that defines the wormhole throat. The invariant characterization of the throat
	that is valid for generic wormholes identifies it as an outer marginal trapped surface subject to additional conditions \cite{HV:98,H:09,MJGMBC:21}, allowing to use our formalism \cite{T:22}.
	
	Consider a dynamical generalisation \cite{SMV:19} of the Simpson--Visser metric \cite{SV:19} that is obtained by rewriting it in the retarded or advanced coordinates $w=u,v$ and allowing the mass paramater to depend it,
	\begin{align}
		ds^2=&-\left(1-\frac{2M(w)}{\sqrt{\eta^2+a^2}}\right)dw^2- 2\epsilon dwd\eta \nonumber \\&+(\eta^2+a^2)d\Omega_2\ . \label{smv}
	\end{align}
	Here $a\geqslant 0$ is a parameter, $-\infty<\eta<+\infty$, and $\epsilon=\pm1$ for $w=u,v$, respectively. As $r=\sqrt{\eta^2+a^2}\geqslant a$, the MSH mass is $C_\mM= 2M+a^2(r-2M)/r^2$.
	
	This metric represents the Schwarzschild spacetime for $a=0$, a regular black or white hole (also known as a black bounce) for $a<2M$, and a traversable wormhole for $a>2M$ \cite{SMV:19}. In the former case, the Schwarzschild radius $r_\epsilon(w)= a\equiv b_0$ marks the wormhole throat, while for a regular black or white hole, $r_\epsilon(w)=2M(w)>a$. Notably, in the static case, the solution falls into the $k=1$ class, with all dynamic cases belonging to $k=0$.
	
	TWH analysis parallels RBH's. Irrespective of quantum gravitational effects during its  formation,   it is presumed that from some stage it is describable by semiclassical gravity.  Finite formation time from a distant observer's perspective and absence of curvature singularities are now part of the basic traversability requirements \cite{V:96}. Ref.~\cite{T:22} demonstrates that dynamical wormholes do not lead to standard Ellis--Morris--Thorne \cite{V:96} or Simpson--Visser \cite{SV:19} static solutions, while allowed static limits are non-traversable and/or violate the bounds on the NEC violation \cite{KS:20}.

	%This metric represents the Schwarzschild spacetime for $a=0$, a spacetime with a regular black or white hole, known also as a black bounce for $a<2M$, and a traversable wormhole for $a>2M$ \cite{SMV:19}. In the former %case the Schwarzschild radius $r_\epsilon(w)= a\equiv b_0$ is located at the wormhole throat, while for a regular black or white hole   $r_\epsilon(w)=2M(w)>a$. It is also easy to see that in the static case the %solution belongs to $k=1$ class, while all dynamical cases correspond to $k=0$.
	
	%The analysis of TWH follows the same lines of analysis as that of RBH. Regardless of what quantum gravitational effects might have played a role in its formation, from come stage it is assumed to be describable by the %semiclassical gravity. Finite time of formation according to a distant observer and absence of the curvature singularities are now also the basic requirements for the wormhole to be traversable. Then Ref.~\cite{T:22} %shows that dynamical wormholes cannot lead to the standard  Ellis--Morris--Thorne \cite{V:96} or Simpson--Visser \cite{SV:19} static solutions, and possible static limits are not traversable and/ or violate bounds %\cite{KS:20} on the amount of violation of the NEC.

	Modification of the construction of Ref.~\cite{SMV:19} allows to bypass this no-go result but faces another obstacle. If the process starts with the metric \eqref{smv} describing a PBH with $r_+(v)=2M(v)>a$, then when the decreasing Schwarzschild radius reaches down to $a$ the object becomes an one-way wormhole \cite{SV:19}. Then, if the mass function $M(v)<a/2$ continues to decrease, the solution describes a TWH with the throat at $b_0=a$. However, if the loss of mass is driven by the Hawking radiation, and  {its} temperature is proportional to $\kappa_\mathrm{K}$, the Hayward--Kodama surface gravity \cite{H:98} (see Supplementary Material for details), then as
	$r_+=2M(v)\to a$ the temperature approaches zero, as $\kappa_\mathrm{K}\to 0$ \cite{MsMT:22}.

	\textit{Discussion}.--- Proving the adequacy of a particular ABH model includes disproving its alternatives. A useful step in this direction is constraining their properties, and the results of this work provide such constraints. PBHs, whether or not they possess an event horizon, do not cosmologically couple, likely from the radiation-dominated era. Theoretically, models that depict RBHs with de Sitter cores must reconcile with the equation of state $p=\rho$ near both the inner and outer horizons.
	
	Additional models for singularity avoidance have been proposed. We demonstrate that black-to-white hole transition aligns with semiclassical gravity, but a macroscopic stable remnant, be it a black or white hole, does not.  %{(The remnant is an extremal configuration-No trapped or anti-trapped region exists.)}.
However, the rate of change in its mass remains unconstrained by these general considerations. Beyond the absence of a known mechanism for producing wormholes, their semiclassical existence is restricted by requiring each mouth to develop from an evaporating black hole.
	
	PBHs exhibit remarkable properties distinguishing them from conventional black holes. Translating these properties into observational signatures will enable answering the question of their role in the physical Universe.
	
	\textit{Acknowledgments}.---   PKD, SM, and IS are supported by an International Macquarie University Research Excellence Scholarship. The work of DRT is supported by the ARC Discovery project Grant No.DP210101279. FS is funded by the ARC Discovery project Grant No. DP210101279.

	\section*{Supplementary Material}
	
	\subsection{Self-consistent approach and physical black holes}	

 To ensure the absence of scalar curvature singularities    we  use  two quantities that can be expressed directly from EMT components,
\begin{align}
	\tilde{\mathrm{T}}\defeq T^\mu_{~\mu}, \qquad \tilde{\mathfrak{T}} \defeq T^\mu_{~\nu} T^\nu_{~\mu} .
\end{align}
The Einstein equations relate them to the curvature scalars as $\tilde{\mathrm{T}} \equiv - {R}/8\pi$ and $\tilde{\mathfrak{T}} \equiv R^\mu_{~\nu} R^\nu_{~\mu}/64\pi^2$. It is possible to show that the component $T^\theta_{~\theta}$ can introduce only sub-leading divergencies, and only the finite values of
\be
     \mathrm{T}=(\tau^r-\tau_t) / f , \qquad \mathfrak{T} =\big( (\tau{^r})^2 + ( {\tau}{_t})^2 - 2 ( {\tau}_t^r)^2 \big) / f^2 ,
     \label{eq:TwoScalars}
\ee
need to be ensured.  From Eqs.~\eqref{split1}--\eqref{eq:mtgEMTdecomp} it follows that the cosmological constant drops out of the consideration of the potentially divergent parts of the effective EMT components.

For $k=0$ solutions the $(tr)$ block of the EMT near the Schwarzschild radius of (an evaporating) PBH is
	\begin{align}
		\ T^a_{~b} = \begin{pmatrix}
			\Upsilon^2/f &   e^{-h}\Upsilon^2/f^2 \vspace{1mm}\\
			- e^h  \Upsilon^2 & -\Upsilon^2/f
		\end{pmatrix},
		\quad
		T_{\hat{a}\hat{b}} = -\frac{\Upsilon^2}{f} \begin{pmatrix}
			1 &1 \vspace{1mm}\\
			1   & 1
		\end{pmatrix},
		\label{tneg}
	\end{align}
	where the second expression is written in an orthonormal frame \cite{MMT:22}.

Using the $(v,r)$ coordinate the metric functions near the (outer) apparent horizon of a PBH can be written as
\begin{align}
& C_+(v,r)=r_+(v)+w_1(v)y+\cO(y^2), \label{mshv}\\
& h_+(v,r)=\chi_1(v) y+\cO(y^2),
\end{align}
where $y\defeq r-r_+$. By its definition $w_1\leqslant 1$. It is exactly zero for Vaidya-like metrics and $w_1=1$ for $k=1$ dynamical solutions that describe the first moment of formation (and, possibly, the last moment of existence) of the trapped region \cite{MMT:22,DSST:23}.

The Hayward--Kodama surface gravity is defined by using the Kodama vector filed $\vKo$ imitating the properties of the Killing vector-based surface gravity \cite{H:98},
\begin{align}
	\frac{1}{2} \vKo^\mu(\nabla_\mu \vKo_\nu-\nabla_\nu \vKo_\mu) \eqdef \kappa_\mathrm{K} \vKo_\nu,
\end{align}
evaluated on the apparent horizon. Hence it is explicitly given by
\begin{align}
	\kappa_\mathrm{K} = \frac{1}{2} \left.\left(\frac{C_+(v,r)}{r^2} - \frac{\partial_r C_+(v,r)}{r}\right) \right|_{r=r_+} \hspace*{-3mm} = \frac{(1-w_1)}{2r_+}. \label{kappaK}
\end{align}
	
	\subsection{Useful cosmological coordinate transformations}

A careful use of relations between different useful coordinate systems is required to study properties of cosmological black holes \cite{vF:15,DSST:23,GV:23}. Here we demonstrate how a metric of Eq.~\eqref{sgenm} with the MSH mass $C_\mM=C+H^2r^3$ (and the metric function given by Eqs.~\eqref{k0met}) can be transformed into a form that makes it obvious that a PBH is embedded in an exponentially expanding flat FLRW universe,
\begin{equation}
ds^2= -d\tau^2+ a^2(\tau) \left(d\chi^2+ \chi^2 d\Omega_2\right)\ . \label{flrws}
\end{equation}

 As the first step we perform a transformation to the Painlev\'{e}--Gullstrand coordinates \cite{FN:98,vF:15,GV:23} via
\be
dt=e^{-h}\left(\mathfrak{A} d\tau+ \mathfrak{B}dr\right)\ ,
\ee
where $\mathfrak{B}=\sqrt{C_\mM(\tau,r)/r}/f$ is chosen to eliminate the pre-factors of $dr^2$, and $\mathfrak{A}$ is the integrating factor,
\be
\pad_r \left(e^{-h}\mathfrak{A}\right)= \pad_\tau \left(e^{-h}\mathfrak{B}\right).
\ee
For transformations of any metric with $C$ and $h$ independent from $t$ (and in particular  pure de Sitter or Schwarzschild--Kottler--de Sitter spacetimes), we set  $dt=d\tau+e^{-h} \mathfrak{B}dr$. We expect $\fA\approx 1$ for $r_\sg\ll r$. The Painlev\'e-Gullstrand form of the metric is
\be
ds^2=-\fA^2d\tau^2+\left(dr-\fA\sqrt{C_\mM/r}d\tau\right)^2+r^2d\Omega_2.
\ee

A the second step the areal radius is expressed as $r=a\chi=a_0e^{H\tau}\chi$ and the metric becomes

\begin{align}
ds^2&= -\fA^2d\tau^2 +a^2\chi^2d\Omega_2 \nonumber \\
&+a^2    \left(d\chi +  \left[H\chi -\fA\sqrt{\frac{C_\mM}{a^3\chi}}\,\right]d\tau\right)^2 .
\end{align}
Taking into account that for $r=a\chi\gg r_\sg$,
\be
\sqrt{\frac{C_\mM}{\chi a^3}}=H\chi-\frac{1}{2Ha}\frac{C}{r^2}+\cO\big(r^{-5}\big),
\ee
and the metric approaches Eq.~\eqref{flrws}

\subsection{Physical white holes}

The spherically symmetric metric with $r_\sg'>0$ describes a regular growing white hole, with timelike anti-trapping horizons bounding the anti-trapped region. In $(u,r)$ coordinates, the line element in the near-horizon region is well-described by
	\be
	ds^2 = -f(u, r)du^2 - 2dudr + r^2d\Omega_2^2.
	\ee
	A singularity is avoided by introducing an inner horizon, arising from the introduction of a minimal length scale.  A useful model is obtained  starting with such a static  {regular white hole}, we promote it to a dynamical one where the evolution will be described by the retarded coordinate $u$. We consider the following metric function
	\be
	f(u, r) = g(u, r)(r - r_-(u))^3(r - r_+(u)),
	\ee
	where $r_+(u)$ is the outer anti-trapping horizon, $r_-(u)$ is the inner anti-trapping horizon, and $g(u, r)$ is a positive function given by
	\begin{align}
		g(u, r)^{-1}&=r^4-3 r_{-} r^3+\left(4 r_{-}^2+3 r_{-} r_{+}\right) r^2\nonumber\\
		&\quad+\left(-3 r_{-}^2 r_{+}-r_{-}^3\right) r+r_{-}^3 r_{+}.
	\end{align}
	This metric was proposed in Ref.~\cite{RFLPV:22} as a means to cure the mass inflation instability problem at the expense of a degenerate inner horizon with vanishing surface gravity. We prescribe an evolution of the inner and outer apparent horizons consistent with the assumptions given above. This implies both horizons must grow, and the two solutions to $r_+(u)=r_-(u)$ describe the formation and disappearance points which occur in an extremal limit \cite{MS:23}.\\
		
   	\subsection{{Black-to-white hole transition}}
   	
  % 	Using the advanced coordinate $v$, the most general spherically symmetric metric is described by the line element
  % 	\begin{align}
  % 	%	ds^2=-e^{h_{+}(v,r)}f(v,r)dv^2+e^{h_{+}(v,r)}dvdr+r^2d\Omega_2.
 %  	\end{align}
  % 	In the literature, the
  Majority of the RBH models exhibit the property of $h_{+}(v,r)=0$, and   we limit our consideration to this particular case. Two key assumptions allow us to make generic statements about the form of the metric function defining a RBH. Firstly, the  trapped region, bounded by the outer apparent horizon $r_{+}(v)$, must be present. Following the definitions of Ref.~\cite{H:94,vF:15} the existence of such a region is equivalent to having $\theta_{-}\theta_{+}>0$, where $\theta_{-}$ and $\theta_{+}$ represent the expansions of ingoing and outgoing rays, respectively. Secondly, ensuring the regularity of the spacetime necessitates the introduction of a minimal length scale, which possibly originates from quantum gravity effects restricted within a finite region. This leads to the presence of an inner horizon $r_{-}(v)$ serving as an additional boundary of the trapped region, satisfying $r_{-}<r_{+}$.
   	
   	Taking into consideration the above properties, we restrict the metric function of the RBH to the form:
   	\begin{align}
   		f(v,r)=g(v,r)(r-r_{-}(v))^{a}(r-r_{+}(v))^b,
   	\end{align}
    where $g(v,r)>0$ is an appropriate function, providing a regular center as well as the desired asymptotic behavior, and $a$, $b$ are positive odd integer numbers, describing the degree of each horizon's degeneracy \cite{MS:23}. According to our analysis, both horizons must decrease in radius and the sole way for the trapped region to vanish is by their merging at a specific time $v_{*}$, i.e. $r_{+}(v_{*})=r_{-}(v_{*})$. At this moment, which is depicted as a gray disk in Fig.~\ref{fig:example}, we have that
    \begin{align}
    	\theta_{-}\theta_{+}|_{v_{*}}=-\frac{2}{r^2}g(v_{*},r)(r-r_{*}(v_{*}))^{a+b}\leqslant0,
    \end{align}
    signifying the disappearance of the trapped region.

   Simultaneously, at this instant of time ($u_{*}=v_{*}$), the transition to a white hole occurs. Following this transition, the dynamical evolution, reminiscent of the RBH case, is described in retarded coordinates to align with the definition of a white hole, as indicated by their expansions. Therefore, the metric function takes the form:
    \begin{align}
    	\tilde{f}(u,r)=\tilde{g}(u,r)(r-\tilde{r}_{-}(u))^{\tilde{a}}(r-\tilde{r}_{+}(u))^{\tilde{b}}.
    \end{align}
    The primary distinction to the RBH case lies in the expanding nature of the anti-trapping horizons. The termination of this expansion is possible in a similar manner to the termination of the RBH evaporation. There exists a moment in time, $u_{\mathrm{f}}$, where these anti-trapping horizons merge, i.e. $\tilde{r}_{+}(u_{\mathrm{f}})=\tilde{r}_{-}(u_{\mathrm{f}})$. Justifying the disappearance of the anti-trapped region is once again based on the sign of the expansions' product: $\theta_{-}\theta_{+}|_{u_{\mathrm{f}}}\leqslant 0$.

	\subsection{{Some properties of the Simpson--Visser metric}}
For the MSH mass $C_\mM(v,r)=2M(v)+a^2\big(r-2m(v)\big)/r^2$ the equation $f(v,r)=r$ has two positive roots $r_1=2M$ and $r_2 =a$. For $M>a/2$ (black hole), the expansion of Eq.~\eqref{mshv} gives
\be
C=2M+\frac{a^2}{4M^2} y+\cO(y^2),
\ee
i.e., $r_+=2M(v)$ and $w_1=a^2/r_+$. For $M<a/2$ (wormhole),
\be
C=a+(4M-a)y/a+cO(y^2),
\ee
i.e., $r_+=a$ and $w_1=4M/a-1$. Eq.~\eqref{kappaK} implies that at the transition $a=2M$ the surface gravity $\kappa_K=0$.
\end{document}